# Optical capillary-based interferometric sensor for detection of gas refractive index

Haijin Chen[1], Xuehao Hu[1,2], Meifan He[1], Pengfei Ren[1], Chao Zhang[3], Hang Qu[1,2,a)]

[1]Research Center for Advanced Optics and Photoelectronics, Department of Physics, College of Science, Shantou University, Shantou, Guangdong 515063, China
[2]Key Laboratory of Intelligent Manufacturing Technology of MOE, Shantou University, Shantou, Guangdong 515063, China
[3]College of Mechanical Engineering, Yangzhou University, Yangzhou, 225127, China
[a)]Corresponding author: haqux@stu.edu.cn

In this paper, we report a capillary-based M-Z interferometer that could be used for precise detection of variations in refractive indices of gaseous samples. This sensing mechanism is quite straightforward. Cladding and core modes of a capillary are simultaneously excited by coupling coherent laser beams to the capillary cladding and core, respectively. Interferogram would be generated as the light transmitted from the core interferes with the light transmitted from the cladding. Variations in refractive index of the air filling the core lead to variations in phase difference between the core and cladding modes, thus shifting the interference fringes. Using a photodiode together with a narrow slit, we could analyze the fringe shifts. The resolution of the sensor was found to be $\sim 1\times 10^{-8}$ RIU, that is comparable to the highest resolution obtained by other interferometric sensors reported in previous literatures. Finally, we also analyze the temperature cross sensitivity of the sensor. The advantages of our sensor include very low cost, high sensitivity, straightforward sensing mechanism, and ease of fabrication.

Recently, R&D of fiber-optic gas refractometer has drawn tremendous attention due to its potential applications in both scientific and industrial fields, such as gas composition detection, gaseous lasers, environmental protection, homeland security, and atmosphere pressure monitoring [1, 2].

Generally, variations in gas refractive index induced by pressure or temperature changes is in a narrow range of $10^{-6}$ to $10^{-4}$ refractive index unit (RIU). Thus, majority of fiber-optic gas refractometers typically adopt interferometric sensing schemes due to their relatively high sensitivities. In the last decade, a number of Fabry-Perot (F-P) or Mach-Zehnder (M-Z) fiber-optic sensors has been proposed and experimentally demonstrated for sensing of gas refractive index. Methods for fabricating F-P fiber sensors may include mechanically immobilizing two single mode fibers (SMFs) in a metallic sleeve while leaving a F-P cavity between the end facets of the two fibers [3], splicing optical capillaries between two SMFs [4,5], splicing hollow-core photonic crystal fibers (HC-PCFs) between two SMFs [6,7], splicing a short piece of SMF between two SMFs with a lateral offset [8], lateral laser-milling an open F-P cavity in the core of a SMF or a multimode fiber [9], creating an air cavity between two fibers via fusion splicing [10], or cascading a SMF with two capillaries (or HC-PCFs) of different inner diameters [11-13]. Spectral interrogations are then performed to correlate the changes in refractive index with spectral shifts in reflection spectra. Sensitivities are typically found to be on the order of $10^3$-$10^4$ nm/RIU. In particular, M. Quan et al. demonstrated a F-P fiber sensor fabricated by cascading a PCF to a capillary fiber fusion-spliced with a section of SMF [12]. This sensor operated based on the Vernier effect and reported an extremely high sensitivity of ~30899 nm/RIU. Assuming that a 1 pm spectral shift could be differentiated (typical resolution of a commercial optical spectrum analyzer), this sensitivity is equivalent to a detection resolution of $\sim 3\times 10^{-8}$ RIU. Though the above-mentioned F-P fiber sensors feature relatively high sensitivities, their fabrications could be technically challenging and costly. Precise fs-laser milling and tricky fiber splicing with meticulous position manipulation may be required, which not only increased complexity of the fabrication process, but also might compromise the robustness of the sensor structure. The use of PCFs and state-of-the-art spectrum analyzers also increased the expense.

M-Z interferometric fiber sensors have also been demonstrated [14-17]. For example, I. Shavrin et al. designed a M-Z interferometer that used a section of SMF as one arm and a section of HC-PCF as the other [16]. Coherent light output from the two fibers generated interferograms captured by a CCD camera. Gaseous samples entering the HC-PCF modified the effective refractive index of the guided mode, thus shifting the interference fringes. Fast Fourier Transform (FFT) was applied to the captured interferograms to analyze the amplitude and phase information associated with the complex refractive indices of gaseous analytes. Finally, a resolution of $10^{-7}$ RIU was reported. Besides, the F-P sensor proposed in Ref. [8] could also be used in a M-Z sensing scheme. By sandwiching a SMF between two SMFs with a lateral offset, the in-coupled light of the lead-in SMF would propagate via the open cavity and the partially-spliced SMF simultaneously, and eventually interfered in the lead-out SMF. Spectral interrogation of the sensor suggested a sensitivity of ~3402 nm/RIU [14]. Note that these sensors also used expensive PCFs or required rather complicated fiber-splicing processes that brought challenges to the sensor fabrication.

Several other fiber-optic gas refractometers utilized in-fiber resonant structures such as fiber gratings (including tilt fiber Bragg gratings or long period gratings) and surface plasmonic resonance (SPR) structures [18-21]. In these sensors, variations in refractive indices at the vicinities of the in-fiber resonant structures would modify the signature wavelength in the transmission or reflection spectra. Sensitivities of these sensors were on the order of $10^3$ nm/RIU, comparable to the interferometric fiber sensors. However, note that creating the resonant structures into fibers generally involved sophisticated phase-mask lithographic techniques and/or multiple depositions of nanometer-thick noble-metal layers.

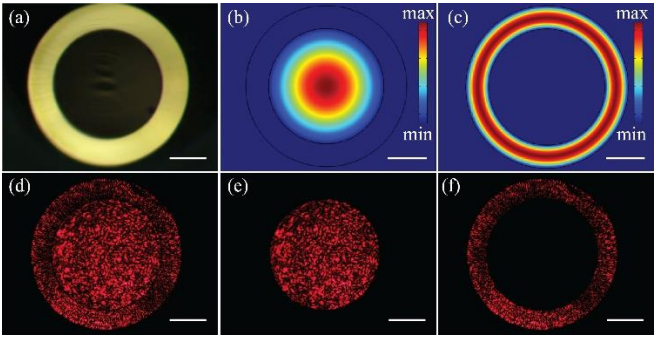

Fig. 1. (a) microscopic image of the cross section of the capillary, (b) fundamental mode guided by the capillary core, (c) fundamental mode guided by the capillary cladding, output intensity of the capillary when laser is coupled to (d) both of the core and cladding, (e) only the hollow core and (f) only the cladding. The scale bar in each image has a length of 100 μm.

In this paper, we propose and experimentally demonstrate an ultrasensitive gas refractometer using an optical capillary-based M-Z interferometer. A coherent laser is split into two virtually-parallel beams that are coupled to the hollow core and cladding of a capillary, exciting core modes and cladding modes thereof. The transmitted light from the cladding and the core of the capillary partially overlaps, generating an interferogram that is then captured by a CMOS camera. Variations in the refractive index of the gaseous samples filling the capillary core could substantially alter the effective refractive index of the core-guided modes, thus shifting the interference fringes. Therefore, sensing of minute changes in the refractive index of the air filling the capillary could be realized by counting the shifted fringes. Moreover, we can also use a photodiode and a slit with a width comparable to that of a single interference stripe (i.e., a bright stripe or a dark stripe) to monitor the amplitude changes in a single-period fringe shift. Experimental results suggest that the sensor proposed has a resolution as high as $10^{-8}$ RIU, which is comparable to the highest resolution reported in previous literatures [12]. Compared to other fiber-optic gas refractometers utilizing complicated sensing structures or requiring tricky fiber processing steps, the sensor proposed in this paper features straightforward sensing mechanism, ultrahigh sensitivity and resolution, as well as low cost. Virtually any commercial optical capillary could be used as the sensing platform in this sensing scheme, and additional fiber processing, such as fs-laser milling or fusion splicing, is not required. The sensitivity of the sensor depends on the length of the capillary, and thus an even higher sensitivity could be achieved by using a longer capillary. The sensor reported here has strong potential for applications such as ultrasensitive gas refractive index detection and precise air-pressure monitoring.

The capillary used in our sensor is a commercial product made from borosilicate glass with a refractive index of ~1.51. the core of the capillary has a diameter of 300 μm and the cladding has a thickness of 60 micron (Fig. 1(a)). Due to the relatively large size of the core and cladding, the in-coupled light thereof would excite a great number of core and cladding modes. A simulation based on Finite Element Method (using COMSOL 5.1) with perfect matched layer (PML) condition is performed to numerically analyze the lowest 50-order modes guided in the core and cladding. In this simulation, we assume that the capillary core is filled with air with a refractive index of 1.0003, and refractive index of the cladding is 1.51 [22, 23]. Simulation results indicate that the effective refractive indices of the lowest 50-order core modes and cladding modes are virtually identical to the indices of the air and borosilicate glass, respectively. Besides, the modal crosstalk between the core and cladding modes is negligible. In Fig. 1 (b, c), we only present the modal distribution of the fundamental modes in the core and cladding. We also experimentally visualize the core modes and cladding modes of the capillary by projecting the output light of the capillary to CMOS camera using a 10× microscopic objective. As shown in Fig. 1(d), both the core and cladding modes are excited, when two individual laser beams are coupled to the core and cladding, respectively. Note that both the core and cladding modes are highly multimoded, thus leading to a scrambled pattern. By physically blocking the beam coupled to the cladding, only the core modes are left, and vice versa (Fig.1(e, f)). We can therefore calculate the cross talk between the core modes and cladding modes by comparing the optical power registered in the cladding region to the power in the core region shown in Fig. 1(e or f). The cross talk is found to be less than -40 dB.

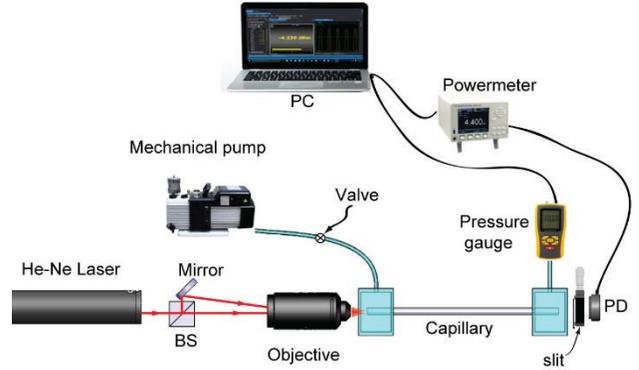

Fig 2. Schematic of the experimental setup of the gas refractometer. BS: beam splitter; PD: photodiode.

The sensing mechanism of the M-Z capillary interferometer could be simply described as follows. By coupling coherent laser beams to the capillary core and cladding, core modes and cladding modes are simultaneously excited and propagate independently. Variations in the refractive index of gaseous samples filling the core would modify the phase difference $\varphi$ of the light propagating in the core and cladding as:

$$\delta\varphi = k_0 l \delta n_{core}, \qquad (1)$$

where $\delta n_{core}$ is the changes in the refractive index of the core; $l$ is the length of the capillary; and $k_0$ is wavenumber defined as $2\pi/\lambda_0$, where $\lambda_0$ is the operating wavelength in vacumm. Experimentally, transmitted light from the core and cladding would partially overlap and produce interference fringes that would shift in response to variations in the index of the gaseous samples filling the core.

Schematic of the proposed capillary sensor is demonstrated in Fig. 2. A He-Ne laser (HNL020LB, Thorlabs) with a wavelength of 632.8 nm and an output power of 1 mW is split into two beams by a cubic beam splitter. Using a 10× microscopic objective, one beam is coupled directly into the hollow core of a borosilicate capillary (length: 10 cm), while the other beam is firstly reflected by a mirror to propagate virtually parallel to its counterpart and then coupled into the capillary cladding by the same objective. The core-guided modes and cladding modes are excited and propagate independently in the capillary waveguide. At the output end, transmitted lights from the core modes and cladding modes interfere, and thus produce an interferogram that is then captured by a CMOS camera. Both ends of the capillary are sealed into the 3D-printed fluidic blocks which enable the air to stream through the capillary rapidly. A transparent window attached on the block enables convenient optical coupling of laser beam into the core and cladding. A mechanical vacuum pump is connected to the block accommodating the capillary input end, and the block holding the other end of the capillary is connected to a digital vacuum gauge with a resolution of 1 hPa ($10^2$ Pa). The correlation between the pressure and refractive index of air can be expressed as [12]:

$$n_{air} = 1 + 7.82 \times 10^{-7} P/(273.6 + T), \quad (2)$$

where $P$ is absolution air pressure (hPa), and $T$ is temperature (°C). Thus, under a constant temperature, refractive index of the air is linearly proportional to the air pressure. We first pump out the air in the capillary to lower the internal air pressure to 400 hPa, and then shut down the mechanical pump to let the air back stream slowly till the inner air pressure elevates to surrounding atmosphere pressure (1012.0 hPa). During this period, the air pressure inside the capillary is continuously logged by a pressure gauge (VC-9200, Lutron Inc.) to synchronize with the fringe shift registered by the camera (Fig. 3).

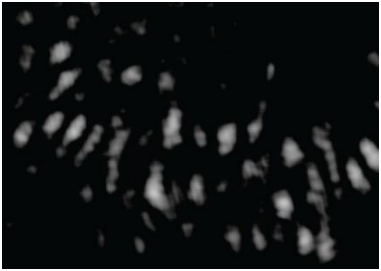

Fig 3. Interferogram partially captured by a COMS camera at the output end of the capillary, while the inner air pressure of the capillary elevates from 400 to 1012 hPa.

We then count the fringe shifts as a function of air pressure and the corresponding refractive index of the capillary core, respectively, as shown in Fig. 4(a, b). Thus, the experimental sensitivity of the sensor could be defined as the number of fringe shifts per refractive index unit (RIU) and is found to be ~$1.55\times10^5$ /RIU. Moreover, we also plot the fringe shifts ($N = \delta\varphi/2\pi$) calculated using Eq. (1,2) (ambient temperature $T$=25 °C). We note that the experimental results agree quite well with theoretical predictions. In Fig. 4(b), we note that the refractive index of the air in the capillary core features an almost perfect linear response to the fringe shifts counted. Thus, assuming a primitive electronic circuit that resolves only a change from the maximum to the minimum of the fringe intensity (0.5 fringe shift) is used, the detection limit of the sensor would be ~$3.22\times10^{-6}$ RIU. In fact, almost two orders of sensitivity improvement could be gained by analyzing full intensity curves as shown in the following discussion.

In a practical air refractive index sensor, instead of using an expensive CCD or COMS camera we would rather use a single photodiode that samples a small area of the interferogram. In our setup, we use a photodiode detector (S150C, Thorlabs) that is placed behind an adjustable slit. To ensure the width of the slit is comparable to that of a single bright (or dark) stripe, we first place the camera behind the slit to visualize the interferogram passing through the slit, and then adjust the width and length accordingly. In our sensor, we set the slit with a length of 1 mm and width of ~200 μm. Then, the camera is replaced by the photodiode probe.

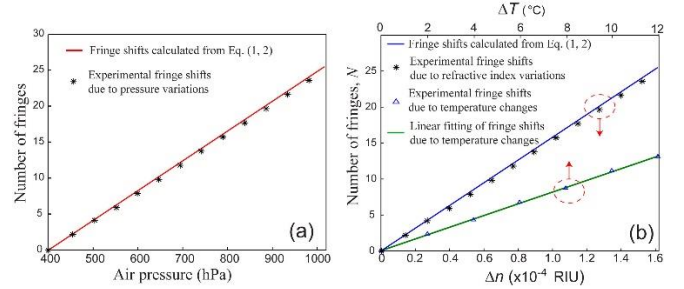

Fig 4. (a) Fringe shifts in response to variations in air pressure in the capillary core, (b) Fringe shifts in response to variations in the refractive index of gas filled in the capillary as well as variations in surrounding temperatures.

In Fig. 5(a-d), we present the recorded intensity variation while the air pressure inside the capillary varies by 50, 100, 150 and 200 hPa. The intensity oscillations in Fig. 5 are a direct consequence of the fringe shifts caused by the changes of the air pressure and thus the refractive index of the capillary core. Therefore, the changes of the refractive index could be quantified by counting the number of the periods in the intensity variation. Note that there is also a good repeatability of the sensor. Particularly, the results in Fig. 5(a-c) look simply like cutouts of Fig. 5(d), although these four measurements are completely independent.

As mentioned earlier, analysis of intensity variations between two fringes, rather than just counting the number of fringe shifts, may provide a much higher sensitivity. For example, in Fig. 5 (b) we could approximately apply a linear fitting between maxima and minima on the second falling edge that corresponds to air pressure changing from 429.8 hPa to 442.2 hPa, and thus a refractive index variation of ~$3.24\times10^{-6}$ RIU according to Eq. (2). Here, the intensity difference between the maxima and the minima is ~30 nW. Assuming that an intensity difference of 100 pW could be reliably detected by a photodiode, we estimate the resolution of our sensor to be ~$1\times10^{-8}$ RIU for gas index sensing. Such a resolution is comparable to the highest resolution that was reported in previous literature [12].

Finally, we would like to comment on the temperature cross sensitivity of the sensor. Perturbations in ambient temperature would lead to changes in the refractive index of the capillary cladding and the air filling the core, as well as minute changes in the diameter and length of capillary, thus resulting in fringe shifts in the interferogram output. The thermo-optic coefficient of borosilicate glass spans a large range depending on the boric content in the glass, so it is challenging to estimate the theoretical temperature cross sensitivity [24]. Experimentally, we investigate the cross sensitivity by heating the capillary sensor using a temperature-controlled breadboard (PTC1, Thorlabs) under ambient atmosphere pressure, and then counting the fringe shifts as a function of temperature increments. As indicated from the linear fitting in Fig. 4(b), the cross sensitivity is ~1 fringe shift per degree (°C). Such a cross sensitivity should be considered, when the capillary sensor is used for high-precision measurements especially for a relatively long-term measurement. On the other hand, our capillary sensor offers almost instantaneous response to variations in gaseous refractive index. In many sensing scenarios that involve quick gas diffusion or pressure variations, changes in gaseous refractive index occur in a short time window so that influence of thermal perturbation on the sensor could be ignored.

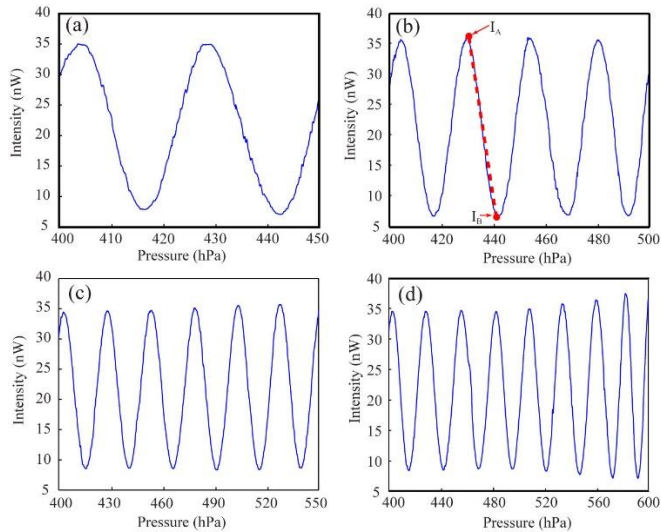

Fig. 5 Intensity variations measured by a photodiode detector behind a slit, when the air pressure in the capillary core changes from 400 hPa to (a) 450 hPa, (b) 500 hPa, (c) 550 hPa, and (d) 600hPa.

In summary, in this paper we demonstrate a capillary-based M-Z interferometer that could be used for precise detection of variations in refractive indices of gaseous samples. The sensing mechanism is quite straightforward. By coupling coherent beams to the capillary cladding and core, the cladding modes and core modes of a capillary are simultaneously excited. The light transmitted from the core interferes with that transmitted from the cladding, thus producing an interferogram at the output end of the sensor. Changes in refractive index of the capillary hollow core lead to variations in the phase difference between the core modes and cladding modes, thus shifting the interference fringes. We interrogate the fringe shifts using a camera or a photodiode with a narrow slit. The resolution of the sensor is found to be $\sim 1\times 10^{-8}$ RIU that is comparable to the highest resolution obtained by other interference sensors reported in previous literatures. The advantages of our sensor include very low cost, high sensitivity, straightforward sensing mechanism and ease of fabrication.

**Acknowledgements**: This project is supported by Department of Education of Guangdong Province (Grant number: 2018KCXTD011).

## References

1. A. Urrutia, I. D. Villar, P. Zubiate and C. R. Zamarreño, Laser Photonics Reviews, 1900094 (2019).
2. J. Hodgkinson and R. P. Tatam, Measurement Science and Technology **24** (1), 012004 (2012).
3. G. Z. Xiao, A. Adnet, Z. Zhang, F. G. Sun and C. P. Grover, Sensors and Actuators A: Physical **118** (2), 177-182 (2005).
4. H. Gao, Y. Jiang, L. Zhang, Y. Cui, Y. Jiang, J. Jia and L. Jiang, Optics express **27** (16), 22181-22189 (2019).
5. M. Hou, F. Zhu, Y. Wang, Y. Wang, C. Liao, S. Liu and P. Lu, Optics express **24** (24), 27890-27898 (2016).
6. F. Yang, Y. Tan, W. Jin, Y. Lin, Y. Qi and H. L. Ho, Optics letters **41** (13), 3025-3028 (2016).
7. J. Tang, G. Yin, C. Liao, S. Liu, Z. Li, X. Zhong, Q. Wang, J. Zhao, K. Yang and Y. Wang, IEEE Photonics Journal **7** (6), 1-7 (2015).
8. D.-w. Duan, Y.-j. Rao and T. Zhu, JOSA B **29** (5), 912-915 (2012).
9. Z. L. Ran, Y. J. Rao, W. J. Liu, X. Liao and K. S. Chiang, Optics express **16** (3), 2252-2263 (2008).
10. C. Liao, T. Hu and D. Wang, Optics express **20** (20), 22813-22818 (2012).
11. R. Wang and X. Qiao, Applied optics **53** (32), 7724-7728 (2014).
12. M. Quan, J. Tian and Y. Yao, Optics letters **40** (21), 4891-4894 (2015).
13. Z. Zhang, J. He, B. Du, K. Guo and Y. Wang, Optics express **27** (21), 29649-29658 (2019).
14. D.-W. Duan, Y.-J. Rao, L.-C. Xu, T. Zhu, D. Wu and J. Yao, Sensors and Actuators B: Chemical **160** (1), 1198-1202 (2011).
15. Z. Li, C. Liao, Y. Wang, L. Xu, D. Wang, X. Dong, S. Liu, Q. Wang, K. Yang and J. Zhou, Optics express **23** (5), 6673-6678 (2015).
16. I. Shavrin, S. Novotny, A. Shevchenko and H. Ludvigsen, Applied Physics Letters **100** (5), 051106 (2012).
17. L. Jiang, L. Zhao, S. Wang, J. Yang and H. Xiao, Optics express **19** (18), 17591-17598 (2011).
18. C. Caucheteur, T. Guo, F. Liu, B.-O. Guan and J. Albert, Nature communications **7**, 13371 (2016).
19. A. K. Mishra, S. K. Mishra and B. D. Gupta, Plasmonics **10** (5), 1071-1076 (2015).
20. T. Allsop, R. Neal, E. M. Davies, C. Mou, P. Bond, S. Rehman, K. Kalli, D. Webb, P. Calverhouse and I. Bennion, Measurement Science and Technology **21** (9), 094029 (2010).
21. F. Tian, J. Kanka and H. Du, Optics Express **20** (19), 20951-20961 (2012).
22. A. Rocha, J. Silva, S. Lima, L. Nunes and L. Andrade, Applied optics **55** (24), 6639-6643 (2016).
23. J. C. Owens, Applied optics **6** (1), 51-59 (1967).
24. J. Ballato and P. Dragic, Materials **7** (6), 4411-4430 (2014).